\documentclass[pra,superscriptaddress,preprintnumbers,showpacs,amsmath,amssymb]{revtex4}
\usepackage{graphicx}

\def\be{\begin{equation}}       \def\ee{\end{equation}}
\def\bea{\begin{eqnarray}}      \def\eea{\end{eqnarray}}

\begin{document}

\title{Exact Calculation of Entanglement in a 19-site 2D Spin System}

\author{Qing Xu}
\affiliation{Department of Physics, Purdue University, West
Lafayette, Indiana 47907, USA}
\author{Sabre Kais\footnote{Corresponding author: kais@purdue.edu}}
\affiliation{Department of Chemistry and Birck Nanotechnology center,
Purdue University, West
Lafayette, Indiana 47907, USA}
\author{Maxim Naumov}
\affiliation{Department of Computer Science, Purdue University, West
Lafayette, Indiana 47907, USA}
\author{Ahmed Sameh}
\affiliation{Department of Computer Science, Purdue University, West
Lafayette, Indiana 47907, USA}

\date{\today}

\begin{abstract}
Using the Trace Minimization Algorithm,
we carried out an exact calculation of entanglement
in a 19-site two-dimensional  transverse Ising model.
This model consists of a  set of localized spin-$\frac{1}{2}$ particles in a two
dimensional triangular lattice coupled through exchange interaction $J$ and
subject to an external magnetic field of strength $h$.
We demonstrate, for such a class of two-dimensional magnetic systems, that
entanglement can be controlled and tuned by varying the
parameter $\lambda=h/J$ in the Hamiltonian and by introducing impurities into the
systems. Examining the derivative of the concurrence as a function of $\lambda$
shows that the system exhibits a quantum phase transition at about $\lambda_c=3.01$,
a transition induced by quantum fluctuations at the absolute zero of temperature.
\end{abstract}


\maketitle

\section{Introduction}

Entanglement, which is  a quantum mechanical property that has no
classical analog, has been viewed as a resource of quantum information
and computation \cite{amico2008,diviccezo,entg1,Nielsen,gruska,vedral,nori1,nori2}.
Intensive researches of entanglement measurement, entanglement monotone,
criteria for distinguishing separable from entangled pure states and all
the extensions from bipartite to multipartite systems have been carried
out both qualitatively and quantitatively \cite{amico2008}. At the interface between quantum
information and statistical mechanics, there has been particular analysis
of entanglement in quantum critical models
\cite{osterloh,osborne,osterloh2004,vidal2003}.

The critical properties in the entanglement allow for a screening of the
qualitative change of the system and a deeper characterization of the
ground state wavefunction undergoing a phase transition. At T=0,
ground states of many-body systems
contain all correlations concerning phases of matters.
Traditionally, systems have been studied by looking, for
example, at their external perturbations, various order
parameters and excitation spectrum \cite{amico2008}. Methods
developed from quantum information shed light on new ways of
studying many-body systems \cite{Lidar1,Lidar2,Lidar3,Lidar4}, such as providing support for numerical
calculations, like density matrix renormalization or design of new
efficient simulation strategies for many-body systems.

Entanglement close to quantum phase transitions was originally analyzed by Osborne
and Nielsen \cite{osborne}, and Osterloh et al. \cite{osterloh}
for the Ising model in one dimension.  Recently, we studied a set of localized
spins coupled through exchange interaction and subject to an external
magnetic filed \cite{Osenda1,Osenda2,Osenda3,Osenda4}. We demonstrated
for such a class of one-dimensional magnetic systems, that
entanglement can be controlled and tuned by varying the
anisotropy parameter in the Hamiltonian and by
introducing impurities into the systems.
In particular, under certain conditions, the entanglement is
zero up to a critical point $\lambda_c$, where a quantum phase transition occurs,
and is different from zero above $\lambda_c$ \cite{kais2002}.

In two and higher dimensions nearly all calculations for spin systems  were
obtained by means of numerical simulations \cite{Sandvik1,Sandvik2}. The concurrence
and localizable entanglement in two-dimensional quantum XY and XXZ models were
considered using quantum Monte Carlo \cite{Syluasen1,Syluasen2}. The results of these calculations
were qualitatively similar to the one-dimensional case, but entanglement is much smaller
in magnitude. Moreover, the maximum in the concurrence occurs at a position closer to the
critical point than in the one-dimensional case \cite{amico2008}.

The Trace Minimization Algorithm for Hermitian eigenvalue problems, like
those under consideration in this paper, was introduced in 1982 by A. Sameh
and J. Wisniewski \cite{sameh1982}. It was designed specifically to handle very large problems
on parallel computing platforms for obtaining the smallest eigenpairs. Later,
a similar algorithm (Jacobi-Davidson) for the same eigenvalue problem was
introduced by Sleijpen and Van der Vorst  in 1996. A comparison of the two
schemes by A. Sameh and Z. Tong in 2000 \cite{sameh2000} showed that the Trace Minimization
scheme is more robust and efficient \cite{Sleijen}.

In this paper, we use the Trace Minimization Algorithm \cite{sameh1982,sameh2000}
to  carry out an exact calculation of entanglement
in a 19-site two dimensional transverse Ising model.
We  classify the ground state properties according to its
entanglement for certain class on two-dimensional magnetic systems
and demonstrate  that
entanglement can be controlled and tuned by varying the
parameter $\lambda=h/J$ in the Hamiltonian and by introducing impurities into the
systems.  The paper is organized as follows.
In the next section, we give a brief overview of the model,
entanglement of formation and the trace minimization algorithm.
Detailed methods are addressed in the appendix. We then proceed
with the results and discussions of 1. the calculation of exact entanglement
of a 19-site spin system, 2. the relationship of entanglement and quantum phase
transition, and 3. the effects of impurities on the
entanglement. The conclusions and the outlook are
presented in the concluding section.

\section{Method}

\subsection{Model}

We consider a set of localized spin-$\frac{1}{2}$ particles in a two
dimensional triangular lattice coupled through exchange interaction $J$ and
subject to an external magnetic field of strength $h$. The
Hamiltonian for such a system is given by
\begin{equation}
H=-\sum_{<i,j>}J_{i,j}\sigma_{i}^x\sigma_{j}^x-h\sum_{i}\sigma_{i}^z,
\end{equation}
where $<i,j>$ is a pair of nearest-neighbors sites on the lattice,
$J_{i,j}=J$ for all sites except the sites nearest to the impurity
site $i_m$, while around the impurity site $J_{i,j}=(1+\alpha)J$, $\alpha$ measures the strength of
the impurity which is located at site $i_m$, and $\sigma_{i}^x$ and
$\sigma_{i}^z$ are the Pauli matrices. For this model it is
convenient to define a dimensionless coupling constant
$\lambda=h/J$.

\subsection{Entanglement of formation}

We confine our interest to the entanglement of two spins, at any
position $i$ and $j$ \cite{osterloh}. We adopt the entanglement of
formation,  a well known  measure of entanglement \cite{wooters1998},
 to quantify our entanglement \cite{kais2002}. All the information
needed in this case is contained in the reduced density matrix
$\rho_{i, j}$. Wootters \cite{wooters1998} has shown, for a pair of binary qubits,
that the concurrence $C$, which goes from $0$ to $1$, can be taken
as a measure of entanglement. The concurrence between sites $i$ and
$j$ is defined as \cite{wooters1998}
\begin{equation}
\label{concurrence}
C(\rho)=max\{0,\epsilon_1-\epsilon_2-\epsilon_3-\epsilon_4\},
\end{equation}
where the $\epsilon_i$'s are the eigenvalues of the Hermitian matrix
$R\equiv\sqrt{\sqrt{\rho}\tilde{\rho}\sqrt{\rho}}$ with
$\tilde{\rho}=(\sigma^y \otimes
\sigma^y)\rho^*(\sigma^y\otimes\sigma^y)$ and $\sigma^y$ is the
Pauli matrix of the spin in y direction.
For a pair of qubits the entanglement can be written as,
\begin{equation}
\label{entanglement}
E(\rho)=\epsilon(C(\rho)),
\end{equation}
where $\epsilon$ is a function of the ``concurrence'' $C$
\begin{equation}
\epsilon(C)=h\left(\frac{1-\sqrt{1-C^2}}{2}\right),
\end{equation}
where $h$ is the binary entropy function
\begin{equation}
h(x)=-x\log_{2}x-(1-x)log_{2}(1-x).
\end{equation}
In this case, the entanglement of formation is given in terms of
another entanglement measure, the concurrence C. The entanglement of
formation varies monotonically with the concurrence.

The matrix elements of the
reduced density matrix needed for calculating the concurrence are
obtained numerically using the formalism developed in the following section.

\subsection{Trace minimization algorithm (Tracemin)}

Diagonalizing a $2^{19}$
by $2^{19}$ Hamiltonian matrix and partially tracing its density matrix
is a numerically difficult task. We propose to compute the  entanglement of formation, first by applying the trace
minimization algorithm (Tracemin) \cite{sameh1982,sameh2000} to obtain the eigenvalues and eigenvectors of the constructed Hamiltonian. Then, we use these eigenpairs and new techniques detailed in the appendix to build partially traced density matrix.

The trace minimization algorithm was developed for computing a few of the smallest
eigenvalues and the corresponding eigenvectors of the large sparse generalized eigenvalue problem
\begin{equation}
AU=BU\Sigma,
\end{equation}
where matrices $A,B \in \mathbb{C}^{n\times n}$ are Hermitian
positive definite, $U=[u_1, ..., u_p] \in
\mathbb{C}^{n\times p}$ and $\Sigma \in \mathbb{R}^{p\times p}$ is a
diagonal matrix. The main idea of Tracemin is that minimizing $Tr(X^* AX)$, subject
to the constraints $X^* BX=I$, is equivalent to finding the
eigenvectors $U$ corresponding to the p smallest eigenvalues. This
consequence of Courant-Fischer Theorem can be restated as
\begin{equation}
\min_{X^* BX=I} Tr(X^* AX)=Tr(U^* AU)=\sum_{i=1}^p\lambda_i,
\end{equation}
where $I$ is the identity matrix. The following steps constitute a
single iteration of the Tracemin algorithm:\\
\begin{tabular}{l l}
$\bullet\quad G=X_{k}^* BX_{k}$ & (compute $G$)\\
$\bullet\quad G=V\Omega V^*$ & (compute the spectral decomposition of $G$)\\
$\bullet\quad H=\tilde{Q}^* A \tilde{Q}$ & (compute $H$, where $\tilde{Q}=X_k V\Omega^{-1/2}$)\\
$\bullet\quad H=W\Omega W^*$ &  (compute the spectral decomposition of $H$)\\
$\bullet\quad \bar{X}_k = \tilde{Q}W$ & (now $\bar{X}_k^*
A\bar{X}_k=\Lambda$ and $\bar{X}_k^* B\bar{X}_k = I$)\\
$\bullet\quad \bar{X}_{k+1}=\bar{X}_k -D$ & ( $D$ is determined s.t. $Tr(X_{k+1}^* AX_{k+1})<Tr(X_k^* AX_k)$ ).\\
\end{tabular}\\
In order to find the optimal update $D$ in the last step, we enforce the
natural constraint $\bar{X}_k^* BD=0$, and obtain
\begin{equation}
\label{tracemin_system_for_D}
\begin{pmatrix}
A & B\bar{X}_k \\
X_k^* B & 0
\end{pmatrix}
\begin{pmatrix}
D\\
L
\end{pmatrix}
=
\begin{pmatrix}
A\bar{X}_k\\
0
\end{pmatrix}
\end{equation}.

Considering the orthogonal projector $P=B\bar{X}_k (X_k^* B^2
X_k)^{-1} \bar{X}_k^*B$ and letting $D=(I-P)\bar{D}$, the linear system \eqref{tracemin_system_for_D} can be rewritten in the following form
\begin{equation}
\label{tracemin_projected_system}
(I-P)A(I-P)\bar{D}=(I-P)A\bar{X}_k.
\end{equation}
Notice that the Conjugate Gradient method can be
used to solve \eqref{tracemin_projected_system}, since it can be shown that the residual and
search directions $\text{r},\text{p} \in \text{Range}(P)^{\perp}$. Also, notice that the linear system \eqref{tracemin_projected_system} need to be solved only to a fixed relative precision at every iteration of Tracemin.

A reduced density matrix, built from the ground state which is obtained by Tracemin,
is usually constructed as follows:
diagonalize the system Hamiltonian $H(\lambda)$, retrieve
the ground state $|\Psi>$ as a function of $\lambda=h/J$, build the
density matrix $\rho=|\Psi><\Psi|$, and trace out contributions of all the other
spins in density matrix to get reduced density matrix
by $\rho(i,j)=\sum_{p}<u_{i}(A)|<v_{p}(B)|\rho|u_{j}(A)>|v_{p}(B)>$,
where ${u_{i}(A)}$ and ${v_{p}(B)}$ are bases of subspaces
$\epsilon(A)$ and $\epsilon(B)$. That includes creating a
$2^{19} \times 2^{19}$ density
matrix $\rho$
followed by permutations of rows, columns and some basic arithmetic
operations on the elements of $\rho$.
Instead of operating on a huge matrix, we pick up only certain
elements from $|\Psi>$, performing basic algebra to build a reduced density matrix
directly. Details are in  the Appendix.

\section{Results and discussions}

\subsection{Exact entanglement of a 19-site spin system}

\begin{figure}
\centering{}
\includegraphics[scale=0.7]{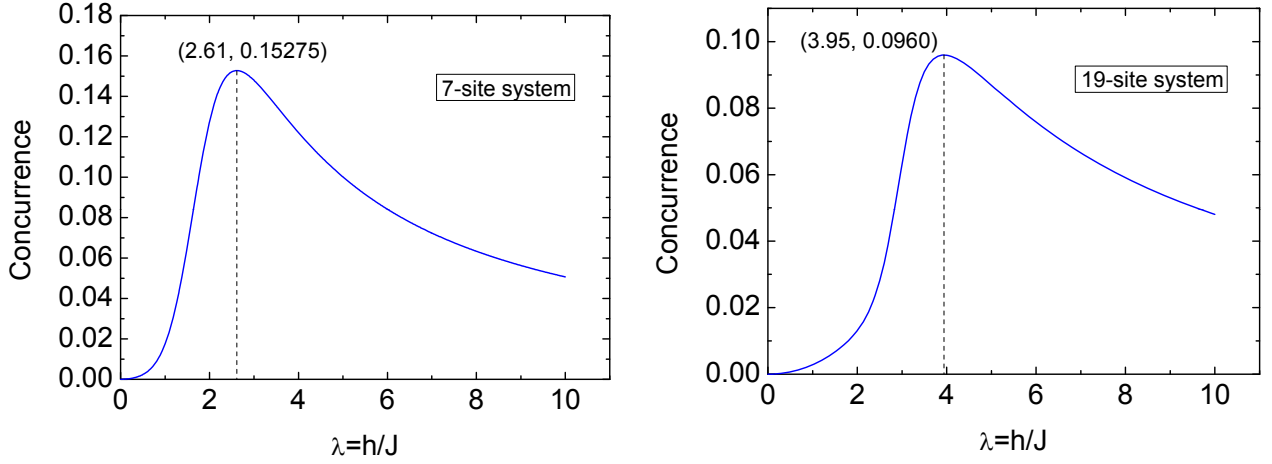}
\caption{(Color online) Concurrence  of center spin and its nearest neighbor as a
function of $\lambda$ for both 7-site and 19-site system.
In the 7-site system, concurrence reaches maximum  0.15275 when $\lambda=2.61$.
In the 19-site system, concurrence reaches the maximum  0.0960 when $\lambda=3.95$. }
\label{fig1}
\end{figure}

We examine the change of concurrence in Eq. \eqref{concurrence} between the center spin and its nearest
neighbor as a function of $\lambda=h/J$ for both the 7-site and 19-site systems.
In Fig.\ref{fig1}, the concurrence of the 7-site system reaches its maximum  0.15275
when $\lambda=2.61$. In the 19-site system, the concurrence reaches
0.0960 when $\lambda=3.95$. The maximum value of concurrence in the 19-site model, where each site interacts with six neighbors, is roughly 1/3 of the maximum concurrence in the one-dimensional transverse Ising model with size N=201 \cite{Osenda1}, where it has only two neighbors for each site. It is the monogamy \cite{Coffman,osborne2006} that limits the entanglement shared among the number of neighboring sites.

\begin{figure}
\centering{}
\includegraphics[scale=0.7]{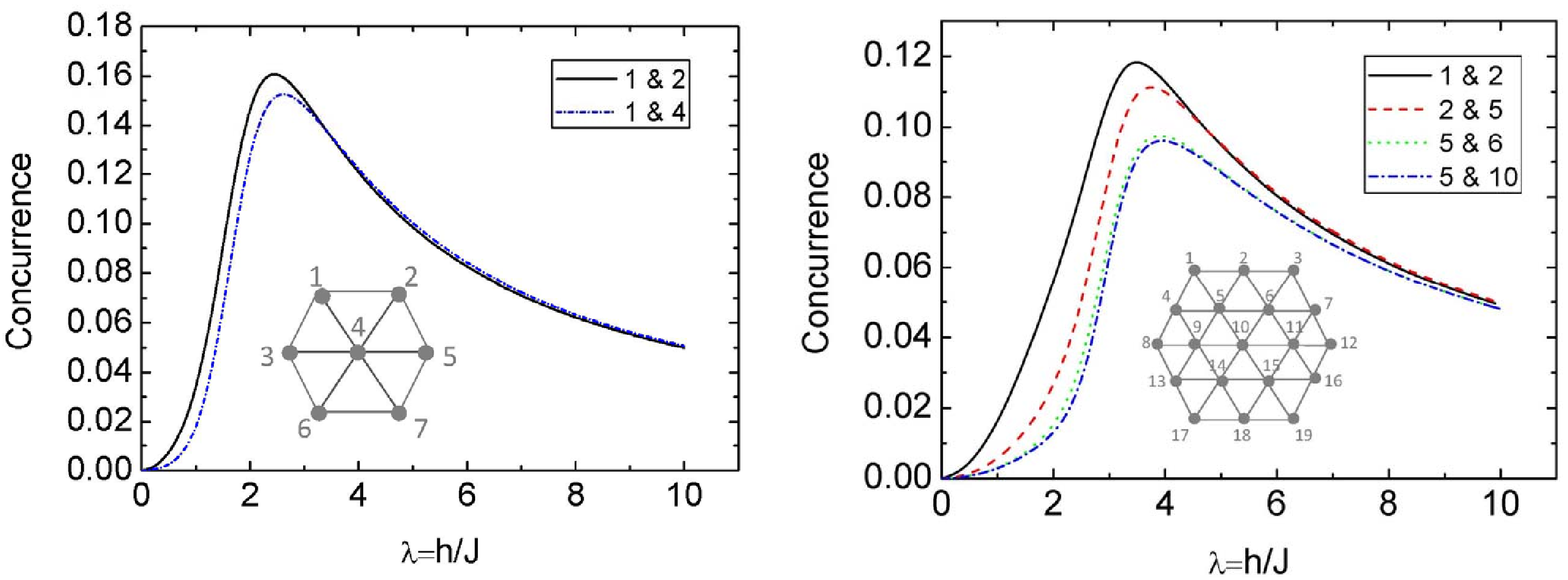}
\caption{(Color online) The nearest neighbor concurrence as a function of $\lambda$ for different pairs. In 7-site system, there are two distinct pairs $1\ \&\ 2$ and $1\ \&\ 4$. In 19-site system, they are $1\ \&\ 2$, $2\ \&\ 5$, $5\ \&\ 6$ and $5\ \&\ 10$. }
\label{fig2}
\end{figure}

However, entanglement between other nearest neighbors are slightly different
than those between the pairs involving the center. Figure \ref{fig2}
shows that the less the number of neighbors of a
pair the larger the entanglement. The concurrence between the 1st and
2nd spins is greater than that between the 1st and 4th in the 7-site system.
For the 19-site one, the concurrence between the 1st and
2nd spins is greater than that between the 2nd and 5th. The same rule applies to
the others, therefore $C_{2,5} > C_{5,6} > C_{5,10}$. Although $5\ \&\ 6$ and $5\ \&\
10$ have the same number of neighbors, the number of neighbors of neighbors of $5\ \&\ 6$
is less than that of $5\ \&\ 10$. Consequently, $C_{5,6}$ is slightly
larger than $C_{5,10}$.


Our numerical calculation shows that the maximum  concurrence of next
nearest neighbor (say the 1st and 10th spins) is less than $10^{-8}$. The
truly non-local quantum part of the two-point correlations is
nonetheless very short-ranged\cite{osterloh}.
It shows that the entanglement is short ranged, though global.
These results are similar to those obtained
for Ising one-dimensional spin systems in a transverse magnetic field\cite{osterloh}.
The range of entanglement, that is the maximum distance between two spins at which
the concurrence is different from zero, is short. The concurrence vanishes unless the two
sites are at most next-neighbours.

\subsection{Entanglement and quantum phase transition}

\begin{figure}
\centering{}
\includegraphics[scale=0.7]{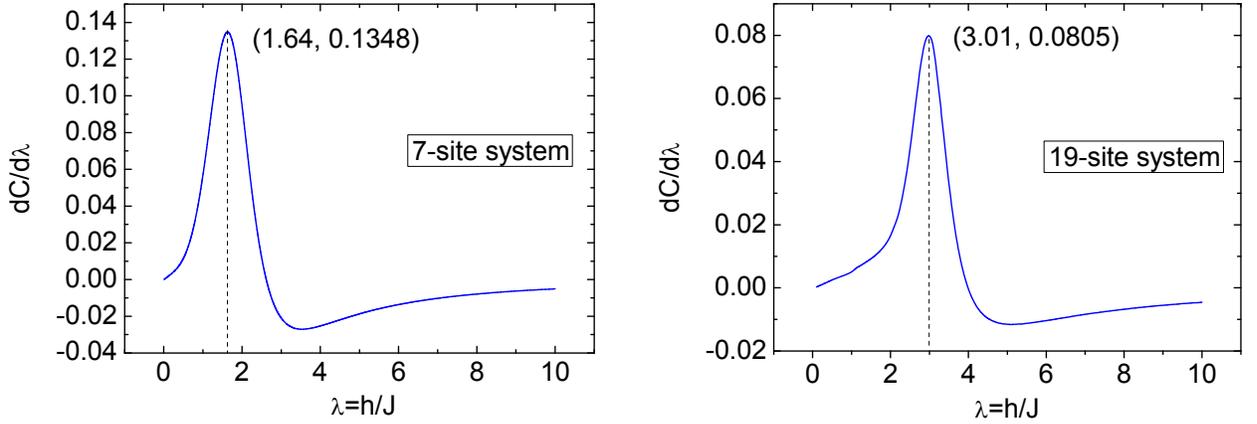}
\caption{(Color online) The change of concurrence between the center and its nearest neighbor as a function of parameter $\lambda$ for 7-site and 19-site systems respectively. They both show the strong tendency of singularity at $\lambda=1.54$ and $\lambda=3.01$.}
\label{fig3}
\end{figure}

As we mentioned in Part II-B, all the information needed for
quantifying the entanglement of two spins is contained in the
reduced density matrix obtained from density matrix. In other words,
entanglement is coded by the information of ground state, while the
quantum phase transition is characterized by the change of ground
state. In order to quantify the change of the many-body wavefunction
when the system crosses the critical point,
we calculate the change of concurrence $dC/d\lambda$ between
the center and its nearest neighbor as a function of parameter $\lambda$ for both 7 and 19 sites
systems, as shown  in Fig.\ref{fig3}. It is known that an
infinite system, a system is at the thermodynamic limit, is supposed to
have a singularity at the critical point of quantum phase transition;
for a finite system one still has to take finite size effect into consideration.
However, in Fig.\ref{fig3} both systems show strong tendency of being singular
at $\lambda_c=1.64$ and $\lambda_c=3.01$ respectively. Renormalization group method for an infinite
triangular system predicts critical point at $\lambda_c=4.75784$; the same method for a square
lattice system at $\lambda_c=2.62975$ \cite{Penson}, while finite-size scaling has $\lambda_c=3.044$
for square lattice \cite{Hamer}. Our results show that the tendency to be singular is moving towards the infinite critical point as the size increases. For one-dimensional system since the calculations can be done for large number of spins, finite size scaling calculations
for N ranging from 41 to 401 spins indicate that the derivative of the concurrence
diverges logarithmically with increasing system size \cite{osterloh}.
In our study we can not perform finite size scaling analysis since we do not have enough
data points to perform data collapse \cite{Kais-Review}. Optimization methods\cite{Kais-Opt} and
Parallel Tracemin code is under development which will allow us to obtain exact results for larger system.

To understand this model better, a discussion about the degeneracy in the system and an explanation
of the energy spectrum is necessary.
It is known that the ground state degeneracy of the Heisenberg spin model depends on whether the total
number of spins is even (singlet) or odd (doublet).
For the Ising model with transverse field on an infinite 1D chain, the ground state in the ferromagnetic
(FM)  phase is doubly degenerate and is gaped from the excitation spectrum by $2J(1-h/J)$ \cite{sachdev}. (Note that,
 however, this degeneracy is never achieved unless one goes to the thermaldynamic limit, regardless
number of spins being even or odd.) In our model, the Ising coupling is ferromagnetic, as opposed to the spin liquid case with
antiferromagnetic coupling, the system is expected to break the $Z_2$ symmetry and develop the (Ising)
FM order under small transverse field. Further, due to the construction of the lattice, it is
impossible to have a system that has even number of sites while conserving all the lattice group
symmetries. So we expect that the same doublet degeneracy remains in 2D as the system goes to the
thermaldynamic limit. The energy gaps from our numerical results of finite systems are less than $10^
{-8}$ (Fig.4), which are well consistent with the expectation. The strict doublets in finite systems only happen at $h/J = 0$ 
exactly, when entanglements naturally are zero, not entangled at all; no matter which one
of the doublet ground state is chosen, it gives the same value of entanglement. Otherwise even very
small h help distinguish the ground state. Technically we don't have to worry about that a different
superposition of the ground states gives different values of entanglement.

\begin{figure}
\centering{}
\includegraphics[scale=0.6]{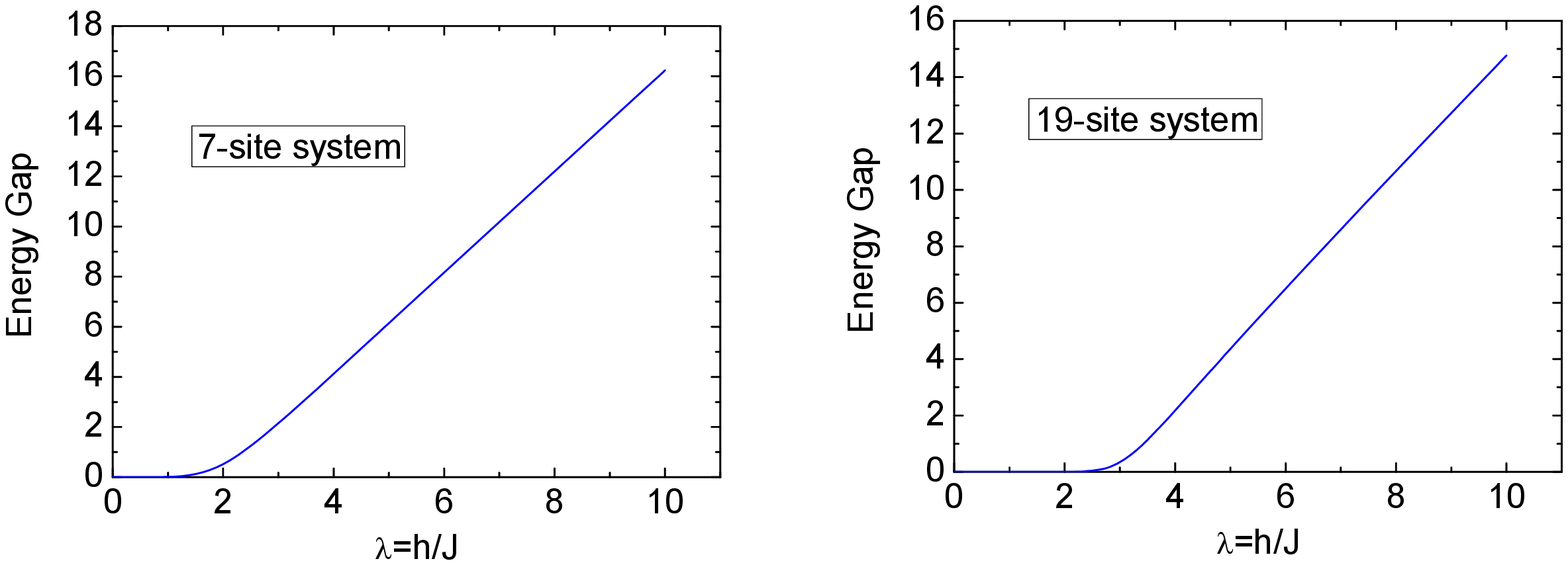}
\caption{(Color online) The energy separation between the ground state and the first
excited state as a function  of $\lambda$.}
\label{fig4}
\end{figure}

The energy separation between the ground state and the first excited state in terms of
$\lambda$ clarifies the spectrum of the system.
Figure \ref{fig4} presents the doubly degeneracy of the ground state and they separate around
$\lambda=1.5$ for the 7-site and $\lambda=2.5$ for the 19-site system. While in the $dC/d\lambda$ vs
$\lambda$ graph, both systems show strong tendency of being singular at $\lambda_c=1.64$ and
$\lambda_c=3.01$ respectively. Both ``separation position'' and ``singular position'' are used
as an indicator of ``critical point''. And we believe using the finite-size scaling both will
give the same ``critical point'' of the infinite system. But it seems $dC/d\lambda$ vs $\lambda$
is a better indicator because for the same size system it points out a value closer to the expected
critical point. This property benefits the finite size scaling method, since less/smaller
systems may be needed.



\subsection{Introducing impurities to tune the entanglement}

\begin{figure}
\centering{}
\includegraphics[scale=0.7]{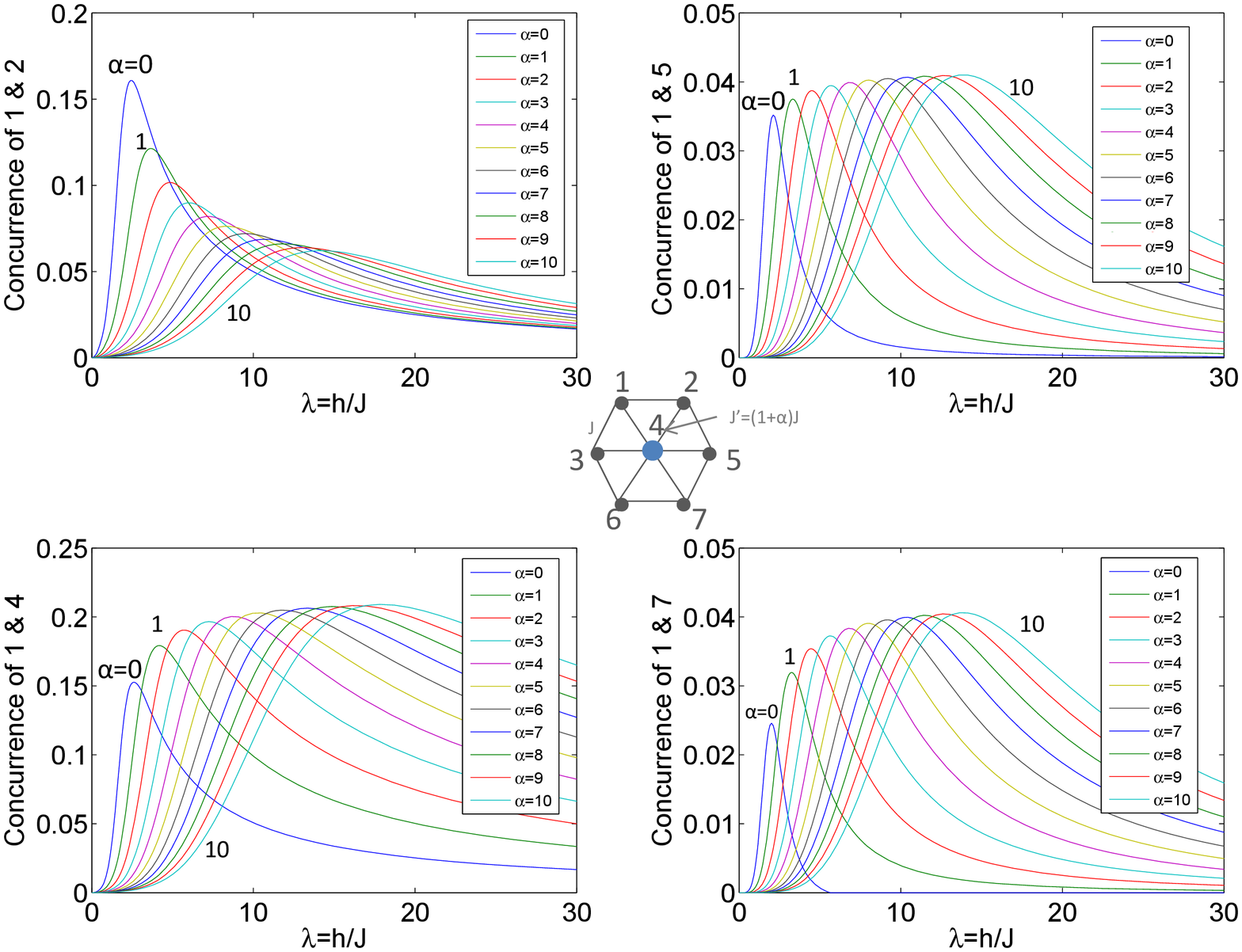}
\caption{(Color online) Concurrences of different pairs with various strength
of impurity in the center of the 7-site system. }
\label{fig5}
\end{figure}

\begin{figure}
\centering{}
\includegraphics[scale=0.7]{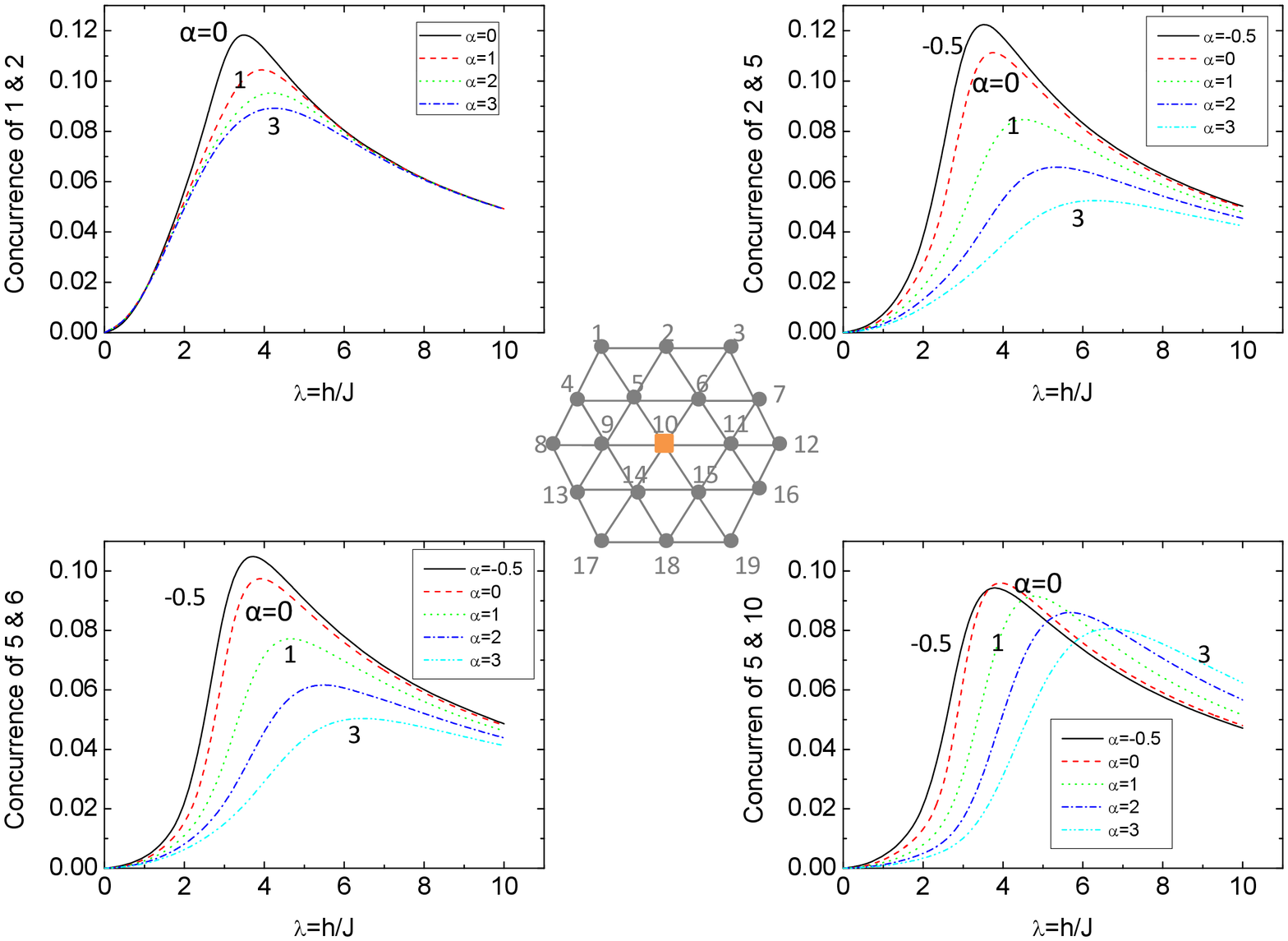}
\caption{(Color online) Concurrence of different pairs with various strength of impurity in the center of the 19-site system. }
\label{fig6}
\end{figure}

\begin{figure}
\centering{}
\includegraphics[scale=0.7]{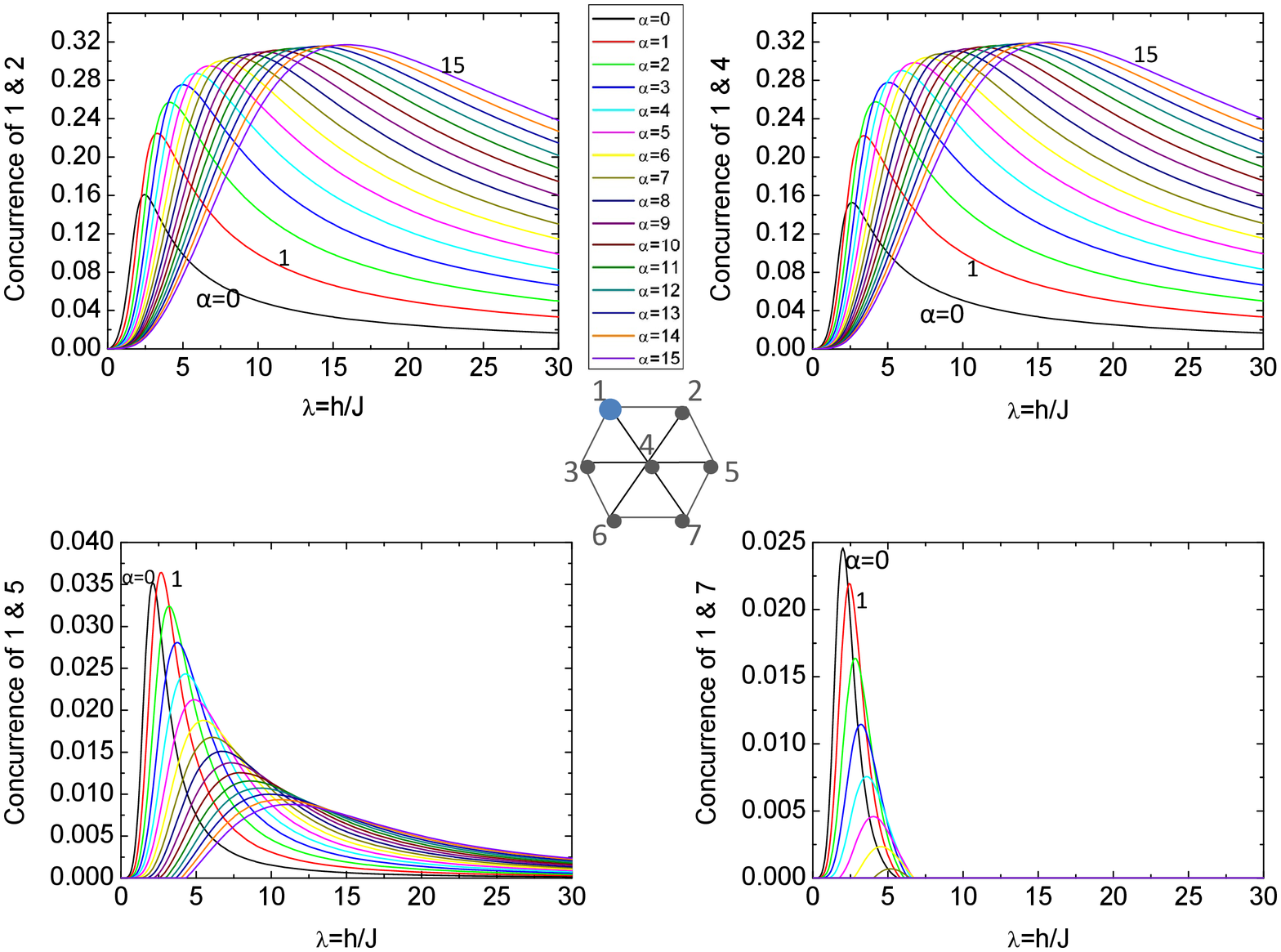}
\caption{(Color online) Concurrence of different pairs with various strength of impurity at site 1 of a 7-site system. }
\label{fig7}
\end{figure}

\begin{figure}
\centering{}
\includegraphics[scale=0.7]{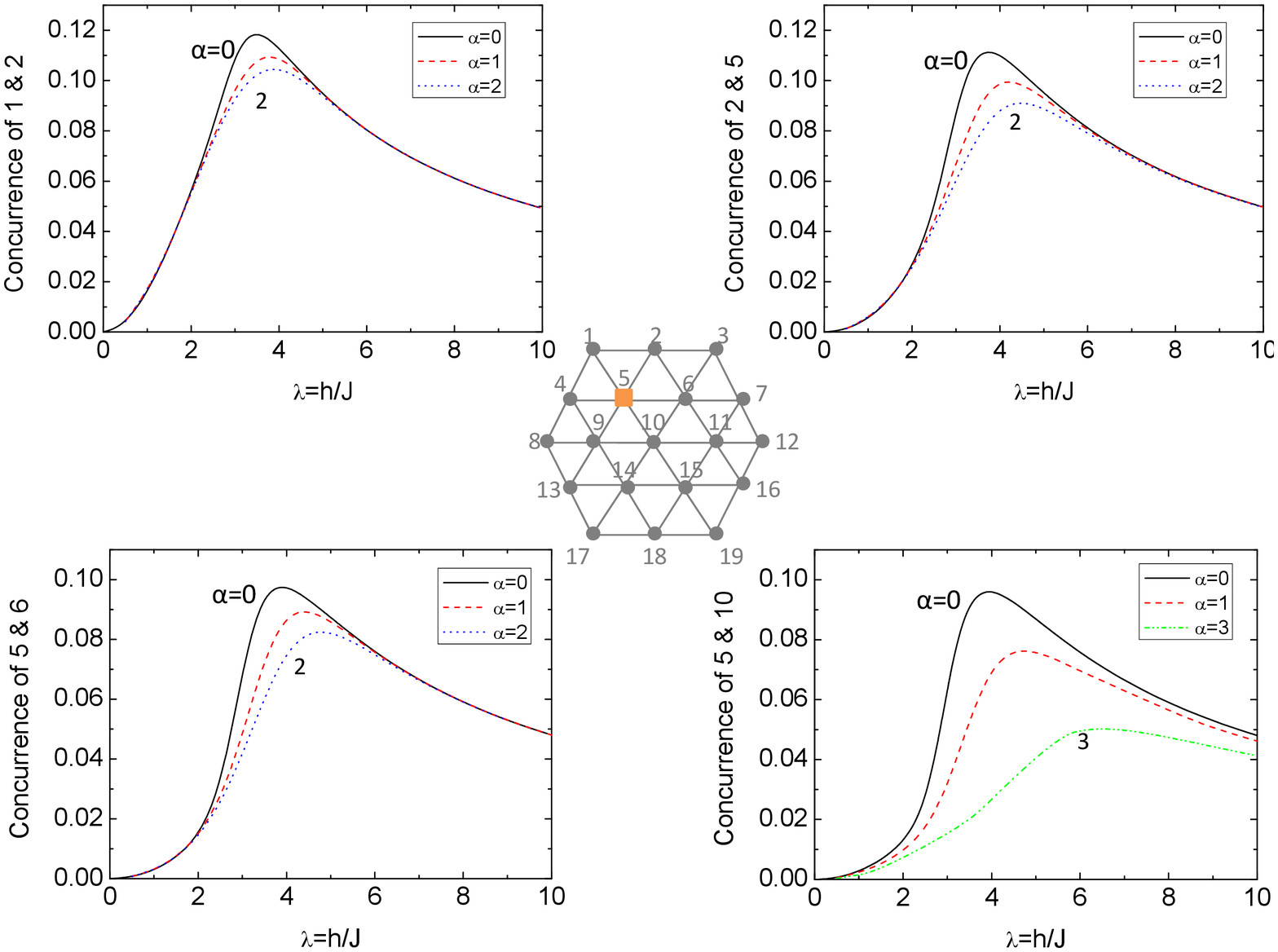}
\caption{(Color online) Concurrence of different pairs with various strength of impurity at site 5 of a 19-site system. }
\label{fig8}
\end{figure}

We introduce one impurity in the center of the 19-site spin system.
The impurity only interacts with nearest neighbors in strength
$J'=(1+\alpha)J$. When the strength increases, the concurrence of any
two spins decreases. Then we move the impurity to site 5. The concurrence shows the same trend of decreasing,
as the impurity strengths go up. Details are shown in Fig.\ref{fig5} for the concurrence of
different pairs with various strength of impurity in the center of the 7-site system, while
Fig.\ref{fig6} show the results for 19-site system.  Fig.\ref{fig7} shows the
results with various strength of impurity at site 1 of a 7-site system and
Fig.\ref{fig8} for various strength of impurity at site 5 of a 19-site system.

The maximum of entanglement is shifted with
the increasing of the parameter $\alpha$. The shift is the result of the competition between the
spin-spin interaction $J$ and the external transverse magnetic field $h$. Consider the ideal
situation of pure infinite system. Without the coupling interaction, all the spins will point
along the direction of transverse field. While with the absence of transverse field, the ground
state is supposed to be 2-fold degenerated, either along the positive x direction or the negative x
direction. Every spin has six neighbors, so averagely is affected by three $J$, and one $h$. When
the two forces are well-matched in strength, the phase transition occurs. Fig. 3 indicates the
19-site system has strong tendency of singularity at $\lambda=3.01$, which is consistent and very
close with the above statement. When the system is finite, the boundary effect (less than six neighbors)
will affect the position of the maximum of entanglement a little bit as Fig. 2 shows for different pairs.
After we introduce the impurity, the balance of $3:1$ is destroyed, so the the maximum of
entanglement is shifted quite a bit for different strength of $J'=(1+\alpha)J$.

The value of the maximum changes with $\alpha$ is because of the monogamy that limits the
entanglement shared among neighbors. For example, in Fig. 4, the stronger the interactions
between 1 $\&$ 4 and 2 $\&$ 4, i.e. the larger $\alpha$, the less 1 $\&$ 2 entangle.
So the value of maximum goes down for larger $\alpha$.

\begin{figure}
\centering{}
\includegraphics[scale=0.7]{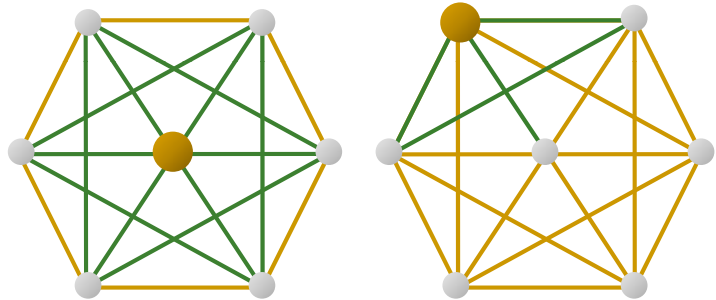}
\caption{(Color online) Overview of the change of concurrence in 7-site system. The large yellow dot stands for the
impurity and silver dots denote regular spins. Lines connecting two sites
represent the entanglement. If the line is green, it means the entanglement
between two sites increases as the impurity gets ``stronger'', otherwise the yellow line indicates that the entanglement decrease when the impurity increases.}
\label{fig9}
\end{figure}

Figure \ref{fig9}
gives a good overview of the change of concurrence for the 7-site system. The large yellow dot stands for the
impurity and silver dots denote regular spins. Lines connecting two sites
represent the entanglement. If the line is green, it means the entanglement
between two sites increases as the impurity gets ``stronger'', and the yellow line indicates that the entanglement decreases when the impurity increases. We can explain these phenomena of 7-site system as follows. When the impurity
interacts more with the neighbor, the pair also entangles more. Since some spins
are more involved with the impurity, they themselves entangle less. The only exceptions are the
next nearest neighbors. Thus,  entanglement close to
the impurity tends to get bigger when J' is greater than J. However, the behavior
of entanglement between site 5 and 10 in the 19-site system surprisingly goes down as
the strength of the impurity coupling increases. It is not clear why the behavior is different
for the one in the 7-site system and whether increasing the system size has any effect.
We are planing to increase the size of the system to include the next layer, which will bring the
system  to 37-site, in order to analyze this phenomena.


All the results above are obtained through sequential computing. In the future to increase the object
size under consideration, we plan to take advantage of parallel computing. We already have a parallel
Tracemin algorithm and we are developing a parallel code for computation of the partial trace. This will
be useful as we expand our 2D systems to larger number of spins in order to perform finite-size
scaling for quantum critical parameters.

In summary, the Tracemin algorithm allowed us
to carry  out an exact calculation of entanglement
in a 19-site two dimensional transverse Ising model.
We demonstrated for such a class of two-dimensional magnetic systems, that
entanglement can be controlled and tuned by varying the
parameter $\lambda$ in the Hamiltonian and by introducing
impurities into the systems.

\section{Acknowledgments}
We would like to thank the Army Research Office (ARO) and the BSF for financial support.

\newpage

\appendix

\section{Applications of trace minimization algorithm}

\subsection{General forms of matrix representation of the Hamiltonian }

By studying the  patterns of $\sum_{<i,j>}I\otimes\cdots
\sigma_{i}^{x}\otimes\cdots \sigma_{j}^{x}\otimes\cdots I$ and
$\sum_{i}I\otimes\cdots \sigma_{i}^z\otimes\cdots I$, one founds the
following rules.

\subsubsection{$\sum_{i} \sigma_{i}^z$ for N spins}

\begin{figure}
\centering
\includegraphics[scale=0.5]{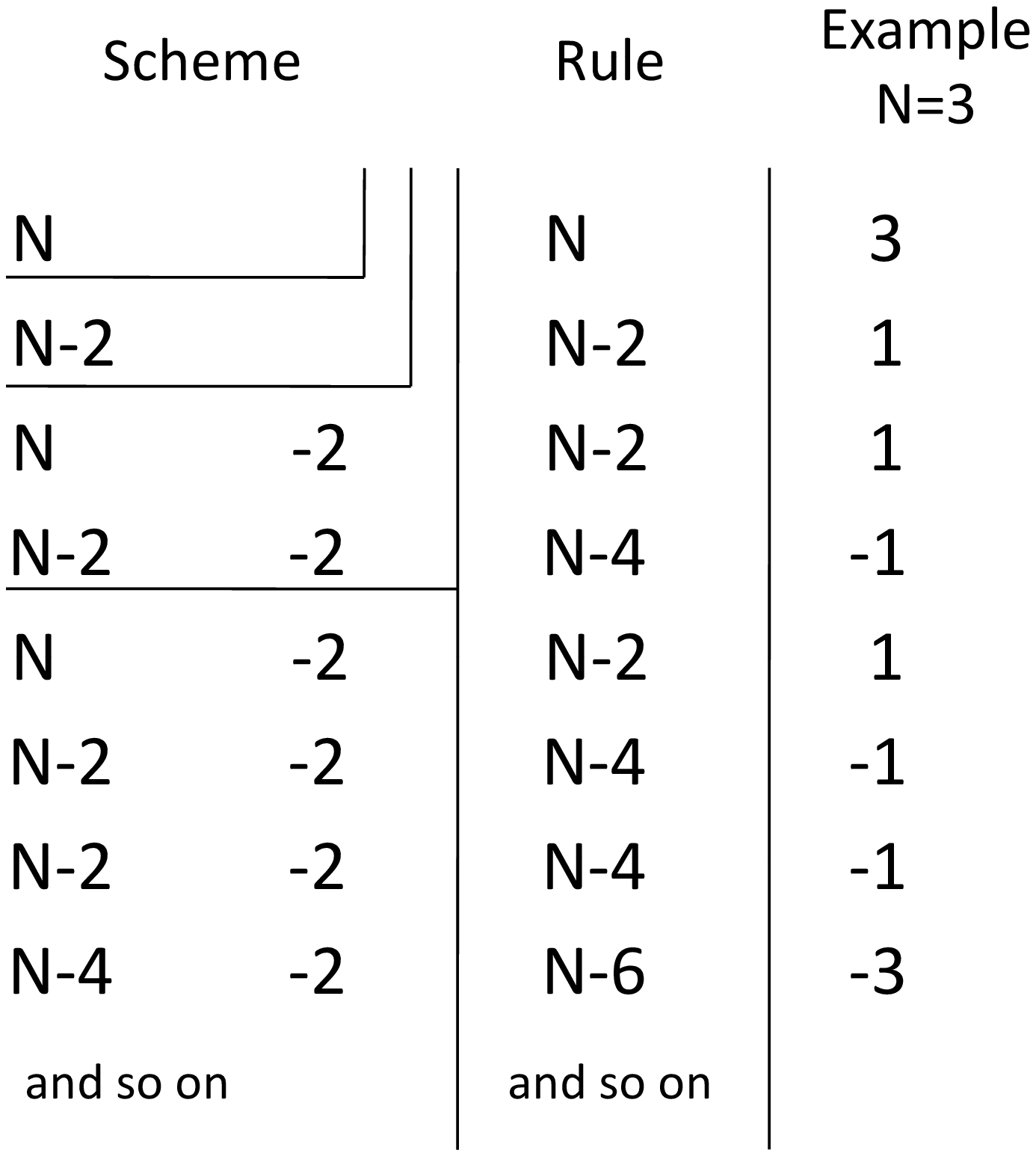}
\caption{Diagonal elements of $\sum_{i} \sigma_{i}^z$ for N spins.}
\label{fig10}
\end{figure}

The matrix is $2^N$ by $2^N$; it has only $2^N$ diagonal elements.
Elements follow the rules shown in Fig. \ref{fig10}.


If one stores these numbers in a vector, and  initializes $v=(N)$,
then the new v is the concatenation of the original v and the original v with 2
subtracted from each of its elements. We repeat this N times, i.e.,
\begin{equation}
v=\left(
\begin{array}{c}
v \\
v-2 \\
\end{array}
\right);
\end{equation}
\begin{eqnarray}
&v&=\left(
\begin{array}{c}
N
\end{array}
\right),\\
\Rightarrow &v&=\left(
\begin{array}{c}
N \\
N-2 \\
\end{array}
\right),\\
\Rightarrow &v&=\left(
\begin{array}{c}
N \\
N-2 \\
N-2 \\
N-4 \\
\end{array}
\right).
\end{eqnarray}

\subsubsection{$\sum_{<i,j>}I\otimes\cdots \sigma_{i}^{x}\otimes\cdots
\sigma_{j}^{x}\otimes\cdots I$ for N spins}

\begin{figure}
\centering{}
\includegraphics[scale=0.4]{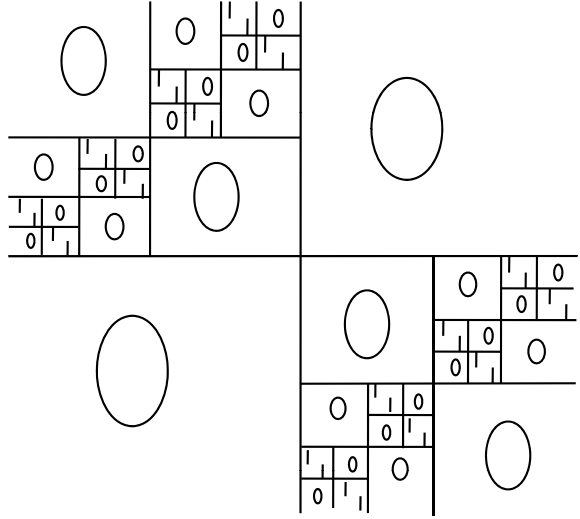}
\caption{Scheme of matrix $I\otimes \sigma_{3}^{x}\otimes
\sigma_{2}^{x}\otimes I\otimes I$}
\label{fig11}
\end{figure}

Since $\left( \begin{array}{ccc}
1 & 0\\
0 & 1\\
\end{array} \right)$ \& $\left( \begin{array}{ccc}
0 & 1\\
1 & 0\\
\end{array} \right)$
exclude each other, for matrix $I\otimes\cdots
\sigma_{i}^{x}\otimes\cdots \sigma_{j}^{x}\otimes\cdots I$, every
row/column contains only one ``1'', then the matrix owns $2^N$
``1''s and only ``1'' in it. If we know the position of ``1''s,
it turns out that we can set a $2^N$ by 1
array ``col'' to store the column position of ``1''s corresponding
to the 1st $\rightarrow$ $2^N$th rows. In fact, the non-zero elements
can be located by the properties stated below. For clarity, let us number N spins in the reverse order
as: N-1, N-2, \dots, 0, instead of 1, 2, \dots, N. The string of non-zero elements starts from the
first row at: $1+2^i+2^j$; with string length $2^j$; and number of such
strings $2^{N-j-1}$. For example, Fig. \ref{fig11} shows these rules  for a scheme of $I\otimes
\sigma_{3}^{x} \otimes \sigma_{2}^{x} \otimes I \otimes I$.


Again, because of the exclusion, the positions of non-zero element
``1'' of $I\otimes\cdots \sigma_{i}^{x}\otimes\cdots
\sigma_{j}^{x}\otimes\cdots I$ are different from those of
$I\otimes\cdots \sigma_{l}^{x}\otimes\cdots
\sigma_{m}^{x}\otimes\cdots I$. So $\sum_{<i,j>}I\otimes\cdots
\sigma_{i}^{x}\otimes\cdots \sigma_{j}^{x}\otimes\cdots I$ is a
$2^N$ by $2^N$ matrix with only 1 and 0.

After storing array ``col'', we repeat the algorithm
for all the nearest pairs $<i,j>$, and concatenate ``col''s to
position matrix ``c'' of $\sum_{<i,j>}I\otimes\cdots
\sigma_{i}^{x}\otimes\cdots \sigma_{j}^{x}\otimes\cdots I$. In the next
section we apply these rules to our problem.

\subsection{Specialized matrix multiplication}
Using the diagonal elements array ``v'' of $\sum_{i} \sigma_{i}^z$ and
position matrix of non-zero elements ``c'' for
$\sum_{<i,j>}I\otimes\cdots \sigma_{i}^{x}\otimes\cdots
\sigma_{j}^{x}\otimes\cdots I$, we can generate matrix H, representing the Hamiltonian. However, we
only need to compute the result of the matrix-vector multiplication H*Y in order to run Tracemin, which is the advantage of Tracemin, and consequently do not need to explicitly obtain H. Since matrix-vector multiplication is repeated many times throughout iterations, we propose an efficient implementation to speedup its computation
specifically for Hamiltonian of Ising model (and XY by
adding one term).

For simplicity, first let Y in H*Y be a vector and $J=h=1$ (in general
Y is a tall matrix and $J\neq h\neq1$). Then
\begin{eqnarray}
&&H*Y \nonumber\\
&=&\sum_{<i,j>}\sigma_i^x \sigma_j^x *Y+\sum_i
\sigma_i^z
*Y\nonumber\\
   &=&\left(
      \begin{array}{cccccc}
        1 &   & 1 &   & 1 &  \\
          & 1 &   & 1 &   & 1\\
          & \ldots&   & \ldots &   &  \\
          &   &   &   &   &  \\
          & \ldots&   &\ldots&   &  \\
        1 &   &   &   & 1 & 1\\
      \end{array}
    \right)*
    \left(
      \begin{array}{c}
        Y(1) \\
        Y(2) \\
        \vdots \\
        \vdots \\
        Y(2^N) \\
      \end{array}
    \right)+
    \left(
      \begin{array}{ccccc}
        v(1)  &   &   &   &   \\
          & v(2)  &   &   &   \\
          &   & \ddots&   &   \\
          &   &   & \ddots&   \\
          &   &   &   & v(2^N)\\
      \end{array}
    \right)*
        \left(
      \begin{array}{c}
        Y(1) \\
        Y(2) \\
        \vdots \\
        \vdots \\
        Y(2^N) \\
      \end{array}
    \right)\nonumber\\
   &=&\left(
      \begin{array}{c}
        Y(c(1,1))+Y(c(1,2))+\ldots+Y(c(1,\#ofpairs) \\
        \vdots \\
        Y(c(k,1))+Y(c(k,2))+\ldots+Y(c(k,\#ofpairs) \\
        \vdots \\
        Y(c(2^N,1))+Y(c(2^N,2))+\ldots+Y(c(2^N,\#ofpairs)\\
      \end{array}
    \right)+
    \left(
      \begin{array}{c}
        v(1)*Y(1) \\
        v(2)*Y(2) \\
        \vdots \\
        \vdots \\
        v(2^N)*Y(2^N)
      \end{array}
    \right),
\end{eqnarray}
where p\# stands for the number of pairs.

\begin{figure}
\centering{}
\includegraphics[scale=0.6]{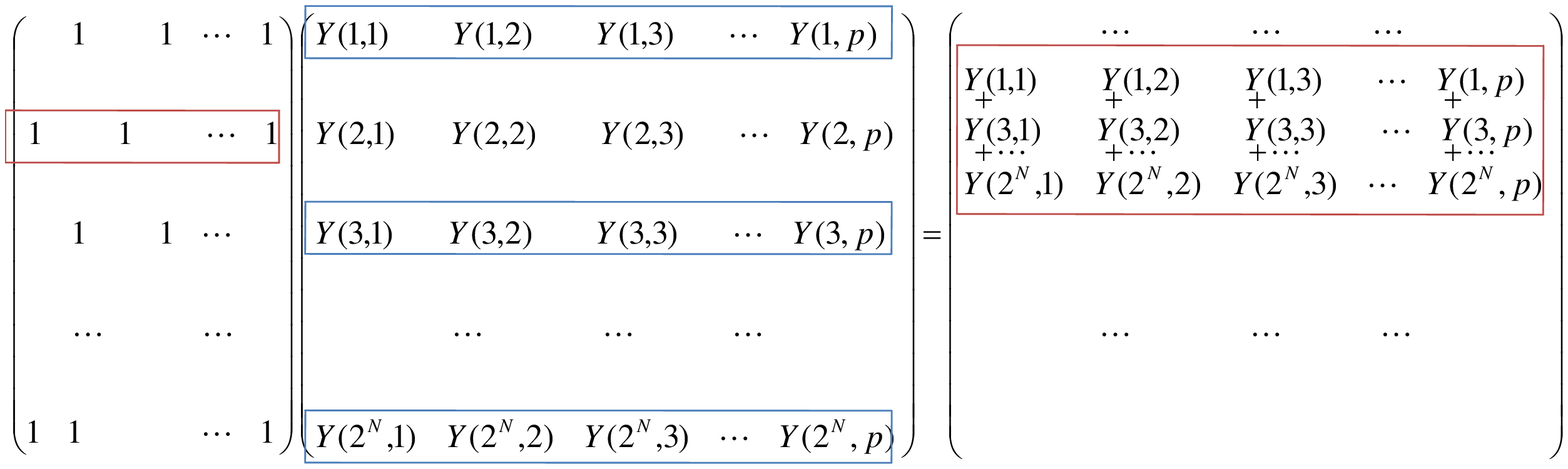}
\caption{(Color online) The illustration of H*Y.}
\label{fig12}
\end{figure}

When Y is a matrix, we can treat Y ($2^N$ by p) column by column
for $\sum_{<i,j>}I\otimes\cdots \sigma_{i}^{x}\otimes\cdots
\sigma_{j}^{x}\otimes\cdots I$. Also, we can accelerate the
computation by treating every row of Y as a vector and adding these vectors at once. Fig. \ref{fig12} visualized the process.


Notice that the result of the multiplication of the xth row of $\sum_{<i,j>}\sigma_i^x \sigma_j^x$ (delineated by the red
box above) and Y, is equivalent
to the sum of rows of Y, whose row numbers are the column indecis of non-zero
elements' of the xth row. Such that we transform a matrix operation to straight forward summation \& multiplication of numbers.

\section{Partial Trace}

All the information needed for quantifying the entanglement of two
spins $i$ \& $j$ is contained in the reduced density matrix
$\rho(i,j)$, which can be obtained from global density matrix
$\rho=|\psi\rangle\langle\psi|$, where $|\psi\rangle$ is the ground state of
the system, via partial trace. Now let us show how we can obtain
the reduced density matrix from the ground state calculated by
Tracemin.

\subsection{Density operator in the  pure case and partial trace}
  Consider a system whose state vector at the instant $t$ is
  \begin{equation}
  |\psi(t)\rangle =\sum_{n}c_{n}(t)|u_{n}\rangle, \quad \sum_{n}|c_{n}(t)|^{2}=1.
  \end{equation}
  The density operator $\rho(t)$ is defined as
  \begin{equation}
  \rho(t)=|\psi(t)\rangle\langle\psi(t)|.
  \end{equation}
  It enables us to obtain all the physical predictions of an observable A(t) by
  \begin{equation}
  \langle A(t)\rangle = Tr\{\rho(t)A\}.
  \end{equation}

  Let us consider two different system (1) and (2) and the global
  system (1)+(2), whose state space is the tensor product:
  $\varepsilon=\varepsilon(1)\otimes\varepsilon(2)$. Let $|u_n(1)\rangle$ be a basis of $\varepsilon(1)$
  and $|v_p(2)\rangle$, a basis of $\varepsilon(2)$, the kets $|u_n(1)\rangle|v_p(2)\rangle$ from a
  basis of $\varepsilon$.

  The density operator $\rho$ of the global system is an operator
  which acts in $\varepsilon$. We construct from $\rho$ an
  operator $\rho(1)$ [or $\rho(2)$] acting only in $\varepsilon(1)$ [or
  $\varepsilon(2)$] which will enable us to make all the physical
  prediction about measurements bearing only on system(1) or
  system(2). This operation will be called a partial trace with
  respect to (2) [or (1)]. Matrix elements of the operator $\rho(1)$ are
  \begin{equation}
  \label{partial_trace}
  \langle u_n(1)|\rho(1)|u_{n'}(1)\rangle=\sum_p(\langle
  u_n(1)|\langle
  v_p(2)|)\rho(|u_{n'}(1)\rangle|v_p(2)\rangle).
  \end{equation}

  Now let A(1) be an observable acting in $\varepsilon(1)$ and $\widetilde{A}(1)=A(1)\otimes
  I(2)$, its extension in $\varepsilon$. We obtain,
  using the definition of the trace and closure relation:
  \begin{equation}
  \widetilde{A}(1)\rangle=Tr\{\rho(1)A(1)\}.
  \end{equation}
  As it is designed, the partial trace $\rho(1)$ enables us to calculate
  all the mean values $\langle \widetilde{A}(1)\rangle$ as if the
  system(1) were isolated and had $\rho(1)$ for a density
  operator \cite{Cohen}.

\subsection{Properties of the reduced density matrix}

  As we calculate the entanglement of formation, we trace out all spins but two
  (bipartite). Their reduced density matrix is, therefore, four by
  four. Reality and parity conservation of H together with translational
  invariance  already fix the structure of $\rho$ to be symmetric
  with  $\rho_{11}$, $\rho_{22}$, $\rho_{33}$, $\rho_{44}$, $\rho_{14}$, $\rho_{23}$
  as the only non-zero entries. It
  follows from the symmetry properties of the Hamiltonian,
  the $\rho$ must be real and symmetrical, plus the global phase
  flip symmetry of Hamiltonian, which implies that $[\sigma_i^z\sigma_j^z,
  \rho_{ij}]=0$, so
  \begin{eqnarray}
  &&
  \left(
    \begin{array}{cccc}
      1 &    &    &  \\
        & -1 &    &  \\
        &    & -1 &  \\
        &    &    & 1 \\
    \end{array}
  \right)
  \left(
    \begin{array}{cccc}
      \rho_{11} &  \rho_{12} &  \rho_{13} & \rho_{14} \\
      \rho_{21} &  \rho_{22} &  \rho_{23} & \rho_{24} \\
      \rho_{31} &  \rho_{32} &  \rho_{33} & \rho_{34} \\
      \rho_{41} &  \rho_{42} &  \rho_{43} & \rho_{44} \\
    \end{array}
  \right)-
  \left(
    \begin{array}{cccc}
      \rho_{11} &  \rho_{12} &  \rho_{13} & \rho_{14} \\
      \rho_{21} &  \rho_{22} &  \rho_{23} & \rho_{24} \\
      \rho_{31} &  \rho_{32} &  \rho_{33} & \rho_{34} \\
      \rho_{41} &  \rho_{42} &  \rho_{43} & \rho_{44} \\
    \end{array}
  \right)
  \left(
    \begin{array}{cccc}
      1 &    &    &  \\
        & -1 &    &  \\
        &    & -1 &  \\
        &    &    & 1 \\
    \end{array}
  \right)=0, \\
  &&
  \left(
    \begin{array}{cccc}
      \rho_{11} &  \rho_{12} &  \rho_{13} & \rho_{14} \\
      -\rho_{21} &  -\rho_{22} &  -\rho_{23} & -\rho_{24} \\
      -\rho_{31} &  -\rho_{32} &  -\rho_{33} & -\rho_{34} \\
      \rho_{41} &  \rho_{42} &  \rho_{43} & \rho_{44} \\
    \end{array}
  \right)-
  \left(
    \begin{array}{cccc}
      \rho_{11} &  -\rho_{12} &  -\rho_{13} & \rho_{14} \\
      \rho_{21} &  -\rho_{22} &  -\rho_{23} & \rho_{24} \\
      \rho_{31} &  -\rho_{32} &  -\rho_{33} & \rho_{34} \\
      \rho_{41} &  -\rho_{42} &  -\rho_{43} & \rho_{44} \\
    \end{array}
  \right)=0, \\
  &&
  2
  \left(
    \begin{array}{cccc}
      0 &  \rho_{12} &  \rho_{13} & 0 \\
      -\rho_{21} &  0 &  0 & -\rho_{24} \\
      -\rho_{31} &  0 &  0 & -\rho_{34} \\
      0 &  \rho_{42} &  \rho_{43} & 0 \\
    \end{array}
  \right)=0, \\
  &&
  \rho_{12}=\rho_{13}=\rho_{21}=\rho_{24}=\rho_{31}=\rho_{34}=0.
  \end{eqnarray}
  Because $\rho_{ij}$ is symmetric,
  \begin{equation}
  \rho_{14}=\rho_{41}, \quad \rho_{23}=\rho_{32},
  \end{equation}
  therefore,
  \begin{equation}
  \rho_{ij}=
  \left(
    \begin{array}{cccc}
      \rho_{11} &  0 &  0 & \rho_{14} \\
      0 &  \rho_{22} &  \rho_{23} & 0 \\
      0 &  \rho_{32} &  \rho_{33} & 0 \\
      \rho_{41} &  0 &  0 & \rho_{44} \\
    \end{array}
  \right).
  \end{equation}

\subsection{Building the reduced density matrix}

  The available code of partial trace involves permutating
  rows/columns of the density matrix $\rho$. On one hand, we need to
  avoid generating the huge matrix $\rho$ ($2^{19}$ by $2^{19}$). On the other hand,
  even if we have $\rho$, permutations are too costly to be computed.
  Fortunately, we are able to
  convert ``generate a global density matrix, then partial trace"
  into ``get six elements then build a reduced density matrix". These six
  elements are closely related to the ground state which we have
  already obtained after the application of Tracemin.
  In fact the structure of the system (the Hamiltonian) guarantees
  that our ground state is a real vector and implicit of time. That
  makes the calculation neater, with no worry about the complex
  conjugate and time evolution.

  Here is how to retreat the six elements. For example, we intent on tracing three spins: 1st, 3rd \& 5th, out of total
  five, provided the ground state with above properties. We start from the definition of partial trace \eqref{partial_trace}
  \begin{eqnarray}
  &&\langle u_n(2)|\langle v_p(4)|\rho(2,4)|u_{n'}(2)\rangle|v_{p'}(4)\rangle\nonumber\\
  &=& \sum_{\alpha,\beta,\gamma}
  \langle a_\alpha(1)|\langle u_n(2)|\langle b_\beta(3)|\langle v_p(4)|\langle c_\gamma(5)|\psi\rangle
  \langle\psi|a_\alpha(1)\rangle |u_{n'}(2)\rangle |b_\beta(3)\rangle |v_{p'}(4)\rangle|c_\gamma(5)\rangle,
  \end{eqnarray}
  and carry it out for a specific matrix element
  \begin{eqnarray}
  &&\rho_{11}(2,4)\nonumber\\
  &=&\langle 00|\rho(2,4)|00\rangle\nonumber\\
  &=&\sum_{\alpha,\beta,\gamma}
     \langle \underline{a_\alpha(1)}\ \underline{0}\ \underline{b_\beta(3)}\ \underline{0}\
     \underline{c_\gamma(5)}|\psi\rangle
     \langle\psi|\underline{a_\alpha(1)}\ \underline{0}\ \underline{b_\beta(3)}\ \underline{0}\
     \underline{c_\gamma(5)}\rangle\nonumber\\
  &&\because|\psi\rangle \text{ is real} \nonumber\\
  &=&\sum_{\alpha,\beta,\gamma}
     \langle \underline{a_\alpha(1)}\ \underline{0}\ \underline{b_\beta(3)}\ \underline{0}\
     \underline{c_\gamma(5)}|\psi\rangle^2.
    \end{eqnarray}
  Since
  \begin{eqnarray}
  &&|\psi\rangle=\sum_{\alpha,n,\beta,p,\gamma}\psi_m|a_\alpha(1)u_{n}(2)b_\beta(3)v_{p}(4)c_\gamma(5)\rangle,\\
  &&\langle a_\alpha(1)u_n(2)b_\beta(3)v_p(4)c_\gamma(5)|\psi\rangle=\psi_m,
  \end{eqnarray}
  $\langle \underline{a_\alpha(1)}\ \underline{0}\ \underline{b_\beta(3)}\ \underline{0}\ \underline{c_\gamma(5)}|\psi\rangle$
  is the coefficient in front of base
  $|\underline{a_\alpha(1)}\ \underline{0}\ \underline{b_\beta(3)}\ \underline{0}\ \underline{c_\gamma(5)}\rangle$
  at $|\psi\rangle$ expansion.
  When we write $|\psi\rangle$ as a column vector $\left(
                      \begin{array}{c}
                        \psi_1 \\
                        \psi_2 \\
                        \vdots \\
                        \psi_{2^N} \\
                      \end{array}
                    \right)$,
  its elements are the coefficients corresponding to the basis $\begin{array}{c}
                                                            |00000\rangle \\
                                                            |00001\rangle \\
                                                            \vdots \\
                                                            |11111\rangle
                                                            \end{array}$.
  Then if we can locate $|\underline{a_\alpha(1)}\ \underline{0}\ \underline{b_\beta(3)}\ \underline{0}\ \underline{c_\gamma(5)}\rangle$
  as the $mth$ base among all basis, we know that $\langle \underline{a_\alpha(1)}\ \underline{0}\ \underline{b_\beta(3)}\ \underline{0}\ \underline{c_\gamma(5)}|\psi\rangle=\psi_m$. Our task is to locate all the basis with the 2nd \& 4th spins being at state
  $|0\rangle$, then pick out corresponding $\psi_m$s, square and sum them together.

\begin{figure}
\centering{}
\includegraphics[scale=0.5]{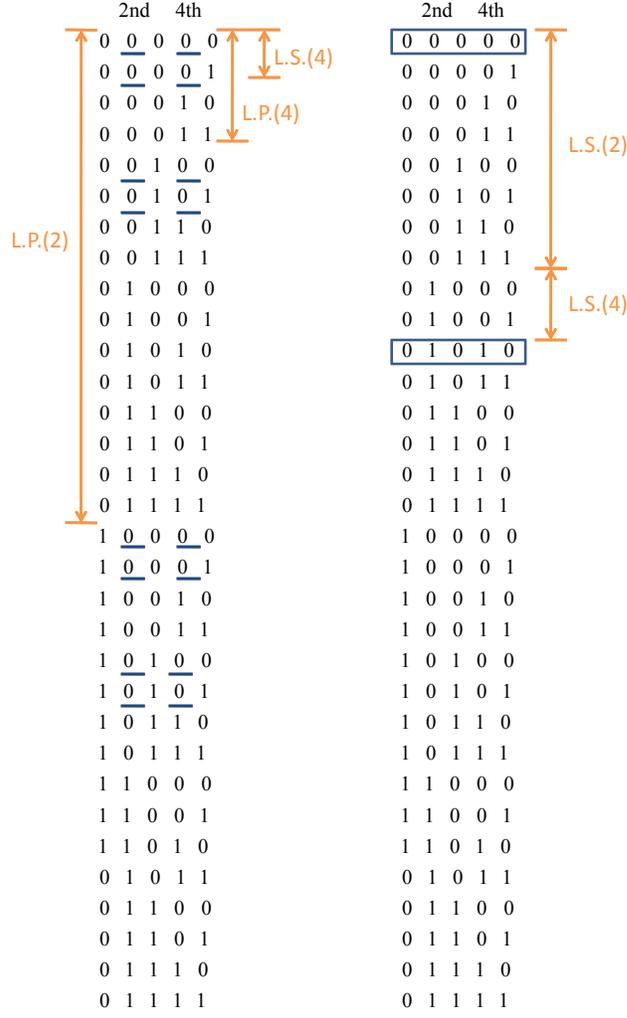}
\caption{(Color online) The illustration of five spins basis.}
\label{fig13}
\end{figure}

  Before we continue onto the details, let us construct a basis matrix of five spins illustrated in Fig.
  \ref{fig13} and define the following.

  $\begin{tabular}{ll}
  \underline{Period}: & we say the pattern like
  $\begin{array}{c}
             0 \\
             \vdots \\
             0 \\
             1 \\
             \vdots \\
             1
           \end{array}$
  is a period.\\
  \underline{Segment}:& $\begin{array}{c}
             0 \\
             \vdots \\
             0
           \end{array}$
  or $\begin{array}{c}
             1 \\
             \vdots \\
             1
           \end{array}$
  is a segment.\\
  \underline{Length}: & the number of elements in a period or a segment.\\
  \underline{L.P.(i)}:& length of a period of the ith spin.\\
  \underline{L.S.(i)}:& length of a segment of the ith spin.
  \end{tabular}$

\vspace{12pt}
The $ith$ spin out of total N spins has:

$\begin{array}{llll}
              & ith/N      & 2nd /5        & 4th/5\\
L.P.          & 2^{N-i+1}  & 2^{5-2+1}=16  & 2^{5-4+1}=4\\
L.S.          & 2^{N-i}    & 2^{5-2}=8     & 2^{5-4}=2\\
\# of periods & 2^{i-1}    & 2^{2-1}=2     & 2^{4-1}=8\\
\# of segments& 2^{i}      & 2^{2}=4       & 2^{4}=16\\
\end{array}
$\\


Using these definitions, we can in general easily locate basis such as
$|\hdots0\hdots0\hdots\rangle$, and within five steps obtain
element $\rho_{11}(i,j)$, by applying the following algorithm.
\begin{enumerate}
\item 1st $0$ in every period for $ith$ spin is located at ``p''= $1,\ 1+L.P.(i),\ 1+2L.P.(i),\hdots\leq2^N$
\item 1st $0$ in every period for $jth$ spin, when the ith is $0$, is located at ``q''= $p,\ p+L.P.(j),\ 1+2L.P.(i),\hdots\leq p+L.P.(i)-1$,
$(\because i<j\ \therefore L.P.(j)<L.P.(i)\ \& L.S.(j)<L.P.(i)).$
\item L.S.(j) decides the length of continued $|\hdots0\hdots0\hdots\rangle$ basis.
\item After locating ``q'', we naturally have $\psi(q)\Rightarrow\psi^2(q)\Rightarrow\psi^2(q)+\psi^2(q+1)+\hdots\psi^2(q+L.S.(j)-1)$,
then locate the next ``q''.
\item When we add them altogether, it is $\rho_{11}(i,j)$.
\end{enumerate}

Similarly, we can locate $``01''\ ``10''\ ``11''$ for
\begin{eqnarray}
\rho_{22}(i,j)&=&\langle 01|\rho(i,j)|01\rangle,\\
  \rho_{33}(i,j)&=&\langle 10|\rho(i,j)|10\rangle,\\
  \rho_{44}(i,j)&=&\langle 11|\rho(i,j)|11\rangle.
\end{eqnarray}
$\rho_{14}(i,j)$ and $\rho_{23}(i,j)$ are a bit different.
\begin{equation}
\rho_{14}(i,j) = \langle00|\rho(i,j)|11\rangle = \langle\hdots0\hdots0\hdots|\psi\rangle\langle\psi|\hdots1\hdots1\hdots\rangle.
\end{equation}
In this case we don't have to locate $|\hdots0\hdots0\hdots\rangle\
\&|\hdots1\hdots1\hdots\rangle$ respectively. Letting $q$ be the position of
$\hdots0\hdots0\hdots$ and the corresponding $q'$ of
$\hdots1\hdots1\hdots$ (i.e. other basis are the same), they are related by
$q'=q+L.S.(i)+L.S.(j)$. That enable us obtain $\rho_{11}$ and $\rho_{14}$
at the same time:
\begin{eqnarray}
\rho_{11}(i,j)&=&\sum\psi^2(q),\\
\rho_{14}(i,j)&=&\sum\psi(q)\psi\left(q+L.S.(i)+L.S.(j)\right).
\end{eqnarray}
So are the $\rho_{22}(i,j)$ and $\rho_{23}(i,j)$.

\newpage

\end{document}